\documentclass[11pt,fullpage]{article}

\usepackage{caption}
\usepackage{subcaption}
\usepackage{amsmath}
\usepackage{xcolor}
\usepackage{colortbl}
\usepackage{natbib}
\usepackage{authblk}
\usepackage{amssymb}
\usepackage{graphicx}
\usepackage{geometry}

\definecolor{forestgreen}{rgb}{0.13, 0.55, 0.13}
\definecolor{Gray}{gray}{0.9}

\begin{document}

\title{Randomization Tests to Assess Covariate Balance When Designing and Analyzing Matched Datasets}

\author[1]{Zach Branson}

\affil[1]{Department of Statistics and Data Science, Carnegie Mellon University}

\maketitle

\begin{abstract}%   <- trailing '%' for backward compatibility of .sty file
Causal analyses for observational studies are often complicated by covariate imbalances among treatment groups, and matching methodologies alleviate this complication by finding subsets of treatment groups that exhibit covariate balance. It is widely agreed upon that covariate balance can serve as evidence that a matched dataset approximates a randomized experiment, but what kind of experiment does a matched dataset approximate? In this work, we develop a randomization test for the hypothesis that a matched dataset approximates a particular experimental design, such as complete randomization, block randomization, or rerandomization. Our test can incorporate any experimental design, and it allows for a graphical display that puts several designs on the same univariate scale, thereby allowing researchers to pinpoint which design---if any---is most appropriate for a matched dataset. After researchers determine a plausible design, we recommend a randomization-based approach for analyzing the matched data, which can incorporate any design and treatment effect estimator. Through simulation, we find that our test can frequently detect violations of randomized assignment that harm inferential results. Furthermore, through simulation and a real application in political science, we find that matched datasets with high levels of covariate balance tend to approximate balance-constrained designs like rerandomization, and analyzing them as such can lead to precise causal analyses. However, assuming a precise design should be proceeded with caution, because it can harm inferential results if there are still substantial biases due to remaining imbalances after matching. Our approach is implemented in the \texttt{randChecks} R package, available on CRAN.
\end{abstract}


\section{Introduction: Matching Approximates Randomized Experiments, But What Kind of Randomized Experiments?} \label{s:introduction}

Randomized experiments are often considered the ``gold standard'' of causal inference because they, on average, achieve balance on all covariates---both observed and unobserved---across treatment groups \citep{rubin2008comment}. However, in observational studies, pretreatment covariates often affect subjects' probability of receiving treatment. As a result, covariate distributions across treatment groups can be very different, leading to biased treatment effect estimates. Furthermore, treatment effect estimates will be more sensitive to model specification, because differing covariate distributions force models to extrapolate; thus, statistical models that make covariate adjustments may still be unreliable \citep{rubin2007design}. Methods must be employed to address this systematic covariate imbalance.

One popular method is matching, which is a prepocessing step that produces a subset of the treatment and control groups that exhibit covariate balance \citep{ho2007matching}. To obtain a subset that exhibits covariate balance, many matching algorithms explicitly pair or block treatment subjects to control subjects that have similar covariate values, as in propensity score matching \citep{rosenbaum1985constructing}, $1:k$ optimal matching \citep{rosenbaum1989optimal,ming2001note,lu2011optimal}, and coarsened exact matching \citep{iacus2011multivariate,iacus2012causal}. Other recent matching algorithms optimize for covariate balance directly, as in covariate-balancing propensity scores \citep{imai2014covariate,zhao2019covariate,vegetabile2019optimally}, kernel optimal matching \citep{kallus2020generalized}, and mixed integer programming approaches \citep{zubizarreta2012using,zubizarreta2014matching,zubizarreta2017optimal}. The main motivation for matching is that it produces a dataset with covariate balance, thereby making treatment effect estimators less biased as well as less sensitive to model misspecification, because there is less need for models to extrapolate when estimating causal effects \citep{ho2007matching,stuart2010matching,abadie2020robust}.

This preprocessing step of finding a subset that exhibits covariate balance is often called the design stage of an observational study, because the goal is to obtain a dataset whose treatment assignment mechanism approximates an experimental design \citep{rubin2007design,rubin2008objective,rosenbaum2010design}. The assumption that a matched dataset approximates a randomized experiment is often justified by demonstrations of covariate balance between the treatment and control groups, as we would expect from a randomized experiment; however, how to best assess covariate balance for matched datasets is still an open problem. Common balance diagnostics are tables and graphical displays of standardized covariate mean differences \citep{ahmed2006heart,stuart2010matching,zubizarreta2012using} and significance tests like $t$-tests and Kolmogorov-Smirnov tests \citep{lu2001matching,bind2019bridging}. For example, a rule-of-thumb is that standardized covariate mean differences of a matched dataset should be below 0.1 \citep{normand2001validating,austin2009some,zubizarreta2012using,resa2016evaluation}. However, many recommend tighter covariate balance if possible: \cite{stuart2010matching} recommends choosing the matching algorithm ``that yields the smallest standardized difference of means across the largest number of covariates,'' and \cite{imai2008misunderstandings} recommends that ``imbalance with respect to observed pretreatment covariates...should be minimized without limit where possible.'' Thus, it is common to run many matching algorithms until some prespecified covariate balance diagnostics are met, and then analyze the matched dataset as if it were from a randomized experiment \citep{dehejia2002propensity,ho2007matching,caliendo2008some,harder2010propensity}.

It is widely agreed upon that covariate balance serves as evidence that a matched dataset approximates a randomized experiment, but what kind of randomized experiment does a matched dataset approximate? Importantly, the aforementioned diagnostics are not formal tests that a matched dataset approximates a randomized experiment; rather, they are rules-of-thumb. For example, because many matching algorithms pair or block treatment subjects with control subjects, it is common to analyze a matched dataset as if it were from a block randomized experiment if covariate balance diagnostics are satisfied, even though these diagnostics do not test the hypothesis that treatment is randomized within blocks \citep{austin2008critical,rubin2008objective,iacus2012causal}. Because of this, it is an ongoing debate as to whether the blocked structure within matched datasets should be ignored when analyzing a matched dataset \citep{ho2007matching,schafer2008average,stuart2010matching,gagnon2019classification}. The choice of assignment mechanism can have substantial implications on inferential results when analyzing a matched dataset, making this an important distinction to clarify. For example, it is well-known that block randomized experiments tend to yield more precise treatment effect estimation than completely randomized experiments (\citealt[Chapter 3]{box1978statistics}, \citealt{greevy2004optimal}, \citealt[Chapter 4]{imbens2015causal}); because of this, \cite{king2019propensity} argued that it is preferable to use a matching algorithm designed to approximate a block randomized experiment (like coarsened exact matching) over a matching algorithm designed to approximate a completely randomized experiment (like propensity score matching). Similarly, recent works have found that rerandomized experiments---where subjects are randomized until covariate balance is achieved---yield more precise treatment effect estimation than completely randomized experiments \citep{li2018asymptotic,li2020rerandomization}. Thus, it is arguably preferable to have a matched dataset that approximates a rerandomized experiment than a completely randomized experiment. Indeed, \cite{stuart2010matching} noted that the iterative approach of matching until covariate balance is achieved is similar to rerandomization, and thus a carefully designed matched dataset may indeed approximate a rerandomized experiment. However, current covariate balance diagnostics cannot ascertain what type of experimental design a matched dataset approximates, if any design at all.

In this work, we develop covariate balance diagnostics that do explicitly test the hypothesis that treatment is randomly assigned, thereby assessing whether or not a matched dataset approximates a completely randomized experiment, a block randomized experiment, or some other experimental design like rerandomization. All of our diagnostics follow a simple two-step procedure: Use a randomization test to quantify the distribution of covariate balance we would expect from a randomized experiment, and then determine if the observed covariate balance in the matched dataset is reasonably within that distribution. Other works have also proposed randomization tests for assessing covariate balance \citep{hansen2008essential,hansen2008covariate,cattaneo2015randomization,lee2013propensity,gagnon2019classification}; our test is a generalization of these procedures, where any experimental design and any measure of covariate balance can be incorporated in the test.

There are four benefits to our procedure. First, our procedure validly tests the hypothesis that treatment is randomly assigned, in the sense that our $\alpha$-level test controls the probability that we falsely reject this hypothesis. Second, our procedure can be easily combined with common diagnostics like Love plots \citep{ahmed2006heart}---which visualize standardized covariate mean differences---thereby providing researchers a data-driven way to assess if the observed covariate balance justifies assuming a particular experimental design for a matched dataset, instead of relying on rules-of-thumb that may vary from researcher to researcher. Third, our procedure allows researchers to assess any experimental design for a matched dataset, including rerandomization designs with covariate balance constraints. This suggests that well-designed matched datasets should perhaps be analyzed as if they are from well-designed rerandomized experiments, a notion we explore throughout this paper. Finally, our procedure allows for a graphical display that puts several experimental designs on the same univariate scale, thereby allowing researchers to pinpoint which experimental design---if any---is most appropriate for a particular matched dataset. All of our tests and diagnostics are provided in the \texttt{randChecks} \texttt{R} package (available on \texttt{CRAN}), such that researchers can easily assess if a treatment is randomly assigned for any dataset.

Our procedure also has limitations, which stem from the fact that our procedure is a balance test---i.e., a procedure that tests for covariate balance. The limitations of balance tests have been widely discussed, to the point that some have recommended against balance tests entirely \citep{imai2008misunderstandings,austin2008critical,stuart2010matching}. The most fundamental limitation of our procedure is that it only tests whether a particular design does \textit{not} hold for a given matched dataset; if our test fails to reject the null hypothesis of random treatment assignment, it may be because the treatment is effectively randomized, or it may be because our test is underpowered in detecting violations of random assignment. This is not a limitation of our test specifically but of hypothesis testing in general. Nonetheless, balance tests are valuable in that they can detect clear violations of random assignment that would undermine causal inferences for matched data \citep{hansen2008essential,lee2013propensity}. Two other critiques of balance tests for matched data---most famously discussed in \cite{imai2008misunderstandings}---are that (1) they often refer to a hypothetical super-population when covariate balance is a characteristic of the matched sample, and (2) their statistical power is affected by sample size. Randomization tests address the first critique by computing an exact randomization distribution for the sample at hand, instead of utilizing asymptotic approximations. However, randomization tests do not address the second critique---indeed, our test will tend to fail to reject random assignment for very small samples due to low statistical power, but it is also the case that analyzing the corresponding matched dataset will have low power when estimating treatment effects \citep{hansen2008essential}. We recommend that researchers quantify the magnitude of covariate imbalances alongside our randomization tests---as is done throughout this paper and in our \texttt{randChecks} \texttt{R} package---so that they can ascertain if covariate balance is still inadequate even if our test fails to reject random treatment assignment.

We recommend the following approach for designing and analyzing matched data. First, researchers should specify the type of experimental design (e.g., complete randomization or block randomization) they would like to approximate from the outset of the observational study. Then, researchers can match subjects in an attempt to approximate this design, where they can use our diagnostics in \texttt{randChecks} to assess if a particular design is plausible. We present these diagnostics in Section \ref{s:randomizationTest}. After an experimental design is deemed plausible, it can be used to analyze the matched dataset; we outline how to do this in Section \ref{s:analysisStage}. In Section \ref{s:simulations}, we will find via simulation that well-designed matched datasets tend to approximate rerandomized experiments with covariate balance constraints, and analyzing matched datasets as such can yield more precise inference for causal effects. However, assuming a precise design should be proceeded with caution, because it can harm inferential results if there are still substantial biases due to covariate imbalances that remain after matching. In Section \ref{s:realDataAnalysis}, we apply our approach to a causal analysis conducted in political science by \cite{keele2017black}, who used subject-matter expertise to target balancing relevant covariates when constructing a matched dataset. We show how our covariate balance diagnostics can ascertain the type of experimental design their matched dataset approximates, as well as how we can condition on this covariate prioritization when analyzing the matched dataset. In Section \ref{s:conclusion}, we conclude with discussions about extending our approach.

\section{The Design Stage: Determining Plausible Experimental Designs for Matched Datasets} \label{s:randomizationTest}

\subsection{Setup, Notation, and Assumptions}

Consider a matched dataset containing $N$ subjects with an $N \times K$ covariate matrix $\mathbf{X}$ and binary treatment assignment $\mathbf{W} \equiv (W_1, \dots, W_N)$, where $W_i = 1$ denotes subject $i$ receiving treatment and $W_i = 0$ denotes control. Let $(Y_i(1), Y_i(0))$ denote subject $i$'s potential outcomes under treatment and control, respectively.\footnote{Such notation implicitly assumes the Stable Unit Treatment Value Assumption \citep{rubin1980comment}, which we discuss shortly.} Only one of the potential outcomes is observed for each subject, depending on the realization of $\mathbf{W}$. We follow the ``Rubin causal model'' \citep{holland1986} and assume that the covariates and potential outcomes are fixed, and thus only $\mathbf{W}$ is stochastic. The treatment and control subjects in the matched dataset may or may not be explicitly paired or blocked. We assume that the causal estimand is the average treatment effect (ATE), $\tau \equiv \frac{1}{N} \sum_{i=1}^N [Y_i(1) - Y_i(0)]$, which may differ from the ATE in some larger, pre-matched dataset. Inference about the $N$ subjects can be generalized to the extent that they are representative of a larger population.

Because only $\mathbf{W}$ is stochastic, it is essential to understand its distribution, which may depend on the potential outcomes $\mathbf{Y}(1) \equiv (Y_1(1),\dots,Y_N(1))$ and $\mathbf{Y}(0) \equiv (Y_1(0),\dots,Y_N(0))$, as well as on the covariates $\mathbf{X}$. However, neither $\mathbf{Y}(1)$ nor $\mathbf{Y}(0)$ are ever completely observed. We employ two assumptions that constrain the distribution of $\mathbf{W}$ to only depend on observed values: Strong Ignorability and the Stable Unit Treatment Value Assumption (SUTVA), which are commonly employed in observational studies \citep{dehejia2002propensity,sekhon2009opiates,stuart2010matching,austin2011introduction}. Strong Ignorability asserts that there is a non-zero probability of each subject receiving treatment, and that---conditional on covariates---the treatment assignment is independent of the potential outcomes \citep{rosenbaum1983central}. SUTVA asserts that the potential outcomes of any subject $i$ depends on $\mathbf{W}$ only through $W_i$ and not other subjects' assignment \citep{rubin1980comment}. Researchers can conduct sensitivity analyses to assess if treatment effect estimates are sensitive to violations of Strong Ignorability (e.g., \citealt[Chapter 4]{rosenbaum2002observational}). See \cite{sobel2006randomized}, \cite{hudgens2008toward}, and \cite{tchetgen2012causal} for a review of methodologies that address SUTVA violations.

\subsection{Formalizing As-If Randomized Assignment in Matched Datasets}

Strong Ignorability implies that $P(\mathbf{W} | \mathbf{Y}(1), \mathbf{Y}(0), \mathbf{X}) = P(\mathbf{W} | \mathbf{X})$, where $P(\cdot)$ denotes the probability measure on the set of all possible treatment assignments $\mathbb{W} \equiv \{0,1\}^N$ \citep{imbens2015causal}. Thus, to conduct causal analyses assuming Strong Ignorability, researchers must assume $\mathbf{W} \sim P(\mathbf{W} | \mathbf{X})$ for some assignment mechanism $P(\mathbf{W} | \mathbf{X})$, i.e., that subjects in the matched dataset are as-if randomized according to a particular assignment mechanism. For example, assuming a matched dataset approximates a completely randomized experiment equates to assuming, for a number of treated subjects $N_T$,
\begin{align}
    \textbf{Complete Randomization}: P(\mathbf{W} = \mathbf{w} | \mathbf{X}) = \begin{cases}
      {N \choose N_T}^{-1} &\mbox{ if } \sum_{i=1}^N w_i = N_T \\
      0 &\mbox{ otherwise.}
    \end{cases} \label{eqn:completeRandomization}
  \end{align}
As another example, assuming a matched dataset approximates a blocked or paired randomized experiment equates to assuming that, for blocks (or pairs) $\mathcal{B}_1,\dots,\mathcal{B}_J$,
\begin{equation}
\begin{aligned}
 \textbf{Block Randomization}: P(\mathbf{W} = \mathbf{w} | \mathbf{X}) &=  \begin{cases}
   \left[ \prod_{j=1}^J {N_j \choose N_{jT}} \right]^{-1} \mbox{ if } \sum_{i \in \mathcal{B}_j} w_i = N_{jT} \hspace{0.05 in} \forall j=1,\dots,J \\
  0 \mbox{ otherwise.}
  \end{cases} \label{eqn:blockRandomization}
\end{aligned}
\end{equation}
where $N_j \equiv |\mathcal{B}_j|$ and $N_{jT}$ denotes the number of treated subjects in $\mathcal{B}_j$.

Complete Randomization (\ref{eqn:completeRandomization}) and Block Randomization (\ref{eqn:blockRandomization}) are commonly assumed when conducting causal inference for matched datasets. For example, (\ref{eqn:blockRandomization}) is commonly assumed after using pairwise or blockwise matching algorithms, such as $1:k$ nearest neighbor matching \citep{rubin1973matching}, $1:k$ optimal matching \citep{rosenbaum1989optimal,ming2001note,lu2011optimal}, full matching \citep{rosenbaum1991characterization,gu1993comparison,hansen2004full,hansen2006optimal}, and coarsened exact matching \citep{iacus2011multivariate,iacus2012causal}. However, some argue that pairwise or blockwise matching is only a conduit to obtain group-level covariate balance, and thus (\ref{eqn:completeRandomization}) can be assumed instead \citep{ho2007matching,schafer2008average,stuart2010matching}. The assignment mechanisms (\ref{eqn:completeRandomization}) and (\ref{eqn:blockRandomization}) are also commonly assumed after using matching algorithms that optimize for group-level covariate balance \citep{zubizarreta2012using,zubizarreta2014matching,kilcioglu2016maximizing,zubizarreta2017optimal}.

As discussed in Section \ref{s:introduction}, researchers typically justify assuming Complete Randomization or Block Randomization by demonstrating that the matched dataset exhibits covariate balance, e.g., standardized covariate mean differences are below 0.1. The most common way to assess covariate balance is to create a Love plot \citep{ahmed2006heart,austin2009balance,zubizarreta2012using}, boxplot \citep{hansen2004full,rosenbaum2012optimal}, or table \citep{dehejia1999causal,harder2010propensity} of standardized covariate mean differences, but other diagnostics like significance tests \citep{smith2005does,lee2013propensity}, graphical displays of propensity score overlap \citep{rubin2000combining,dehejia2002propensity,crump2009dealing,li2018addressing}, and machine learning metrics \citep{linden2016using,gagnon2019classification} are also common. Thus, researchers will often iteratively match subjects until covariate balance diagnostics are satisfactory, but there has also been a recent surge in algorithms that ensure balance diagnostics are satisfactory by design in one step instead of iterative steps \citep{iacus2012causal,zubizarreta2012using,imai2014covariate,vegetabile2019optimally}. However, neither Complete Randomization nor Block Randomization fully conditions on the balance criteria that modern matching algorithms are designed to provide, whether it be iteratively or in one step. For example, the following assignment mechanism (which we call \textit{Constrained Randomization}) conditions on the standardized covariate mean differences being below 0.1:
\begin{align}
    \textbf{Constrained Randomization}: P(\mathbf{W} = \mathbf{w} | \mathbf{X}) &= \begin{cases}
        |\mathcal{A}|^{-1} &\mbox{ if } \mathbf{w} \in \mathcal{A} \\
        0 &\mbox{ otherwise.} \label{eqn:constrainedRandomization}
    \end{cases}
\end{align}
where $\mathcal{A} \equiv \{\mathbf{w}: |\bar{\mathbf{x}}_T - \bar{\mathbf{x}}_C | < 0.1 \text{ and } \sum_{i=1}^N w_i = N_T\}$ denotes the set of \textit{constrained randomizations}, and $\bar{\mathbf{x}}_T - \bar{\mathbf{x}}_C$ denotes the standardized covariate mean differences between treatment and control. Constrained Randomization is similar to rerandomization, where subjects are randomized until covariate balance is achieved \citep{morgan2012rerandomization,li2018asymptotic,branson211ridge}. To our knowledge, Constrained Randomization has not been used in the design and analysis of matched data, and we will explore its use throughout this paper.

How do we know when it is appropriate to assume Complete Randomization, Block Randomization, Constrained Randomization, or some other assignment mechanism for a matched dataset? While there are many covariate balance diagnostics and many matching algorithms designed to satisfy those diagnostics, current diagnostics do not assess which assignment mechanism should be assumed for a matched dataset. In Section \ref{ss:randomizationTest}, we present a randomization test for the hypothesis that treatment within a matched dataset follows a particular assignment mechanism (e.g., Complete Randomization, Block Randomization, or Constrained Randomization). In addition to appraising the covariate balance of a matched dataset, our randomization test procedure allows researchers to determine which assignment mechanism---if any---is most appropriate for a matched dataset, as we show in Section \ref{ss:graphicalDiagnostic}.

\subsection{A Test for As-If Randomized Assignment in Matched Data} \label{ss:randomizationTest}

Here we outline a randomization test for the hypothesis that $H_0: \mathbf{W} \sim P^*(\mathbf{W} | \mathbf{X})$ in a matched dataset. We use the notation $P^*(\mathbf{W} | \mathbf{X})$ to denote that this is a distribution posited by the researcher, rather than the \textit{true} distribution of treatment assignment, which is never known in an observational study. The intuition behind our test is that it computes the distribution of balance we would expect if we conducted a randomized experiment on the data at hand using $P^*(\mathbf{W} | \mathbf{X})$ as the assignment mechanism. If the observed balance is substantially within the distribution of balance we would expect from a particular experimental design, then we do not find evidence against assuming that design. The test is as follows.  \\

\noindent\fbox{%
\parbox{\textwidth}{%
\textbf{$\alpha$-level Randomization Test for $H_0: \mathbf{W} \sim P^*(\mathbf{W} | \mathbf{X})$ in a Matched Dataset}
\begin{enumerate}
\setlength\itemsep{0em}
  \item Specify an assignment mechanism $P^*(\mathbf{W} | \mathbf{X})$, which defines $H_0$.
  \item Define a test statistic $t(\mathbf{W}, \mathbf{X})$, which measures covariate balance.
  \item Generate random draws $\mathbf{w}^{(1)},\dots,\mathbf{w}^{(M)} \sim P^*(\mathbf{W} | \mathbf{X})$, where $M$ is reasonably large (e.g., 1,000) to approximate the randomization distribution.
  \item Compute the randomization distribution of covariate balance:
  \begin{align}
    \left(t(\mathbf{w}^{(1)}, \mathbf{X}), \dots, t(\mathbf{w}^{(M)}, \mathbf{X})\right) \label{eqn:balanceDistribution}
  \end{align}
  \item Compute the following randomization-based two-sided $p$-value:
  \begin{align}
    p = \frac{1 + \sum_{m=1}^M \mathbb{I}(|t(\mathbf{w}^{(m)}, \mathbf{X})| \geq |t^{obs}|)}{M + 1}, \hspace{0.1 in} \text{where $t^{obs} \equiv t \left(\mathbf{W}^{obs}, \mathbf{X} \right)$} \label{eqn:pvalue}
  \end{align}
  \item Reject $H_0$ if $p \leq \alpha$.
\end{enumerate}
  }
  } \\

\noindent
An immediate benefit of the above randomization test is that it is a valid and exact test for the null hypothesis that a specific assignment mechanism holds within a population of matched subjects; this readily follows from classical results on the validity of randomization tests \citep{edgington2007randomization,good2013permutation,hennessy2016conditional,branson2019randomization}. Thus, our randomization test can be used to validly assess the plausibility of a given experimental design: If one rejects the null hypothesis $H_0$ for an assignment mechanism $P^*(\mathbf{W} | \mathbf{X})$, then that mechanism is not appropraite for the matched dataset. As discussed in Section \ref{s:introduction}, failing to reject the null hypothesis does not ``prove'' that random assignment holds, but it nonetheless serves as evidence that assuming a particular assignment mechanism for a matched dataset may be appropriate. Within the test, we recommend setting $\alpha = 0.15$, following common recommendations of other balance tests \citep{cattaneo2015randomization,hartman2018equivalence}; we elaborate on this point at the end of this section.

Our randomization test requires the researcher to specify a test statistic $t(\mathbf{W}, \mathbf{X})$ that measures covariate balance. Importantly, $t(\mathbf{W}, \mathbf{X})$ is not a function of the outcomes, which prevents researchers from biasing results when designing the matched dataset \citep{rubin2007design,rubin2008objective}. The power of our randomization test depends on the relevance of the test statistic; see \cite{rosenbaum2002observational} and \cite{imbens2015causal} for discussions of test statistic choices for randomization tests. We will focus on the standardized covariate mean differences \citep{stuart2010matching,zubizarreta2012using} and Mahalanobis distance \citep{mahalanobis1936generalized,rosenbaum1985constructing, gu1993comparison, diamond2013genetic}, because they are the most commonly used measures for appraising balance in matched datasets. Setting $t(\mathbf{W}, \mathbf{X}) = |\bar{\mathbf{x}}_T - \bar{\mathbf{x}}_C |$ (i.e., using the standardized covariate mean differences as a test statistic) results in a randomization test $p$-value for each covariate, thereby allowing for covariate-by-covariate assessments. Meanwhile, setting $t(\mathbf{W}, \mathbf{X})$ equal to the Mahalanobis distance results in a single $p$-value, thereby allowing for a global assessment of covariate balance. The Mahalanobis distance is defined as\footnote{Here, $\text{cov}(\mathbf{\bar{x}}_{T} - \mathbf{\bar{x}}_{C}) = \frac{N}{N_T N_C} \text{cov}(\mathbf{X})$, where $\text{cov}(\mathbf{X})$ is the sample covariance matrix of $\mathbf{X}$, which is fixed across randomizations. This equality is derived in \cite{morgan2012rerandomization}.}
\begin{align}
    M &= (\mathbf{\bar{x}}_{T} - \mathbf{\bar{x}}_{C})^T \left[ \text{cov}(\mathbf{\bar{x}}_{T} - \mathbf{\bar{x}}_{C}) \right]^{-1}(\mathbf{\bar{x}}_{T} - \mathbf{\bar{x}}_{C}), \label{eqn:md}
  \end{align}
In Section \ref{ss:graphicalDiagnostic} we demonstrate how to create easy-to-interpret graphical displays of our test using the standardized covariate mean differences or the Mahalanobis distance. In general we recommend using the Mahalanobis distance as a test statistic, because it accounts for the joint behavior of covariates, instead of only assessing marginal balance for each covariate.

Our randomization test also requires generating random draws $\mathbf{w}^{(1)},\dots,\mathbf{w}^{(M)} \sim P^*(\mathbf{W} | \mathbf{X})$. Sometimes these draws can be generated via permutations of the observed treatment assignment $\mathbf{W}^{obs}$; this is the case for generating draws from Complete Randomization in (\ref{eqn:completeRandomization}) and Block Randomization in (\ref{eqn:blockRandomization}). In other cases, these draws can be generated via rejection-sampling: For example, to generate draws from Constrained Randomization in (\ref{eqn:constrainedRandomization}), one can generate draws from Complete Randomization in (\ref{eqn:completeRandomization}) and only accept a draw if $|\bar{\mathbf{x}}_T - \bar{\mathbf{x}}_C | < 0.1$. If rejection-sampling is computationally intensive, importance-sampling can be used to approximate the randomization test $p$-value \citep{branson2019randomization}.

Our test is a generalization of other  balance tests. For example, \cite{hansen2008covariate} and \cite{hansen2008essential} proposed permutation tests using the Mahalanobis distance as a test statistic, and \cite{gagnon2019classification} proposed a permutation test using machine-learning methods to construct a test statistic. Another test is the Cross-Match test \citep{rosenbaum2005exact}, which focuses on the pairwise nature of matched datasets. Permutation tests have also been used to assess the balance of subjects in regression discontinuity designs \citep{cattaneo2015randomization,matteiMealli2016}. All of these tests are special cases of our randomization test, where draws from $P^*(\mathbf{W} | \mathbf{X})$ correspond to permutations of $\mathbf{W}^{obs}$.

However, as noted elsewhere in the literature (e.g., \cite{cattaneo2015randomization} and \cite{hartman2018equivalence}), Type II errors are a concern for balance tests like ours, because we want to avoid falsely concluding that treatment is effectively randomized when really it is not. One option for avoiding Type II errors is to set $\alpha$ to be larger than 0.05 (e.g., \cite{cattaneo2015randomization} recommend setting $\alpha = 0.15$, and \cite{hartman2018equivalence} noted that many researchers choose 0.15 or 0.2). \cite{hartman2018equivalence} recommend using equivalence tests instead of balance tests, which essentially ``flip the null and alternative,'' such that the null hypothesis is that the data are not randomized and the alternative is that they are randomized, thereby avoiding the issues of a test being underpowered. The equivalence tests of \cite{hartman2018equivalence} are a promising way to assess randomized treatment assignment, but \cite{hartman2018equivalence} focused only on covariate-by-covariate assessments of Complete Randomization, and it is unclear how to extend their approach to other assignment mechanisms or omnibus metrics like the Mahalanobis distance.

Relatedly, the power of our randomization test depends on---in addition to the test statistic---the alternative hypothesis that holds, if not the randomized assignment null hypothesis $H_0: \mathbf{W} \sim P^*(\mathbf{W} | \mathbf{X})$. Thus, studying the power of balance tests like ours requires quantifying the discrepancy between random assignment and the true assignment mechanism. For example, \cite{hansen2009propensity} quantified the discrepancy between propensity score matched data and block randomization using the within-block variability of propensity scores and covariates, and established that the power of balance tests like ours tends to one if this variability is asymptotically large (e.g., Proposition 2.1 of \cite{hansen2009propensity}). An interesting avenue for future work is characterizing the discrepancy between other designs---such as Constrained Randomization---and alternative assignment mechanisms that may hold in matched data, thereby providing a way to study the power of balance tests like ours.

\subsection{Graphical Diagnostics for Assessing Different Designs: The Lalonde Data} \label{ss:graphicalDiagnostic}

As an example of how to use our randomization test---and how graphical diagnostics can complement our test---we will consider the Lalonde dataset \citep{lalonde1986evaluating}, which has been extensively used for evaluating matching methods \citep{dehejia1999causal,dehejia2002propensity,smith2005does,iacus2012causal,diamond2013genetic}. The data are available in the \texttt{MatchIt} \texttt{R} package \citep{stuart2011matchit}; the treatment group consists of 185 individuals who participated in the National Supported Work Demonstration, and the control group consists of 429 individuals sampled from the Population Survey of Income Dynamics. There are eight covariates: age, years of education; whether someone is black, whether someone is Hispanic, whether or not someone is married, whether or not someone has a high school degree, and income in 1974 and 1975. We defer to the aforementioned references for fuller descriptions of the data, since they have been extensively studied previously. The tests and visuals in this section are reproduced as examples in our \texttt{randChecks} \texttt{R} package; all the code used to conduct these tests and create these visuals is provided in Appendix \ref{s:appendixA}.

We will assess the balance of three datasets: The full Lalonde data, a dataset from 1:1 propensity score matching, and a dataset from 1:1 cardinality matching. We implemented 1:1 propensity score matching using the \texttt{MatchIt} \texttt{R} package and a logistic regression to estimate the propensity scores. Meanwhile, we implemented cardinality matching using the \texttt{designmatch} \texttt{R} package \citep{zubizarreta2016designmatch}, which finds the largest subset of the observational data that fulfills prespecified covariate balance constraints \citep{zubizarreta2014matching}. In our implementation of cardinality matching, we required the eight standardized covariate mean differences to be less than 0.1. In Sections \ref{s:simulations} and \ref{s:realDataAnalysis} we focus on cardinality matching, because it guarantees that covariate balance constraints hold within a matched dataset. However, cardinality matching may discard treated subjects to achieve balance constraints: The 1:1 propensity score matched dataset contained 370 subjects, while the 1:1 cardinality matched dataset contained 240 subjects.

\begin{figure}
\centering
\begin{subfigure}[t]{0.45\textwidth}
  \centering
  \includegraphics[width=1.1\textwidth]{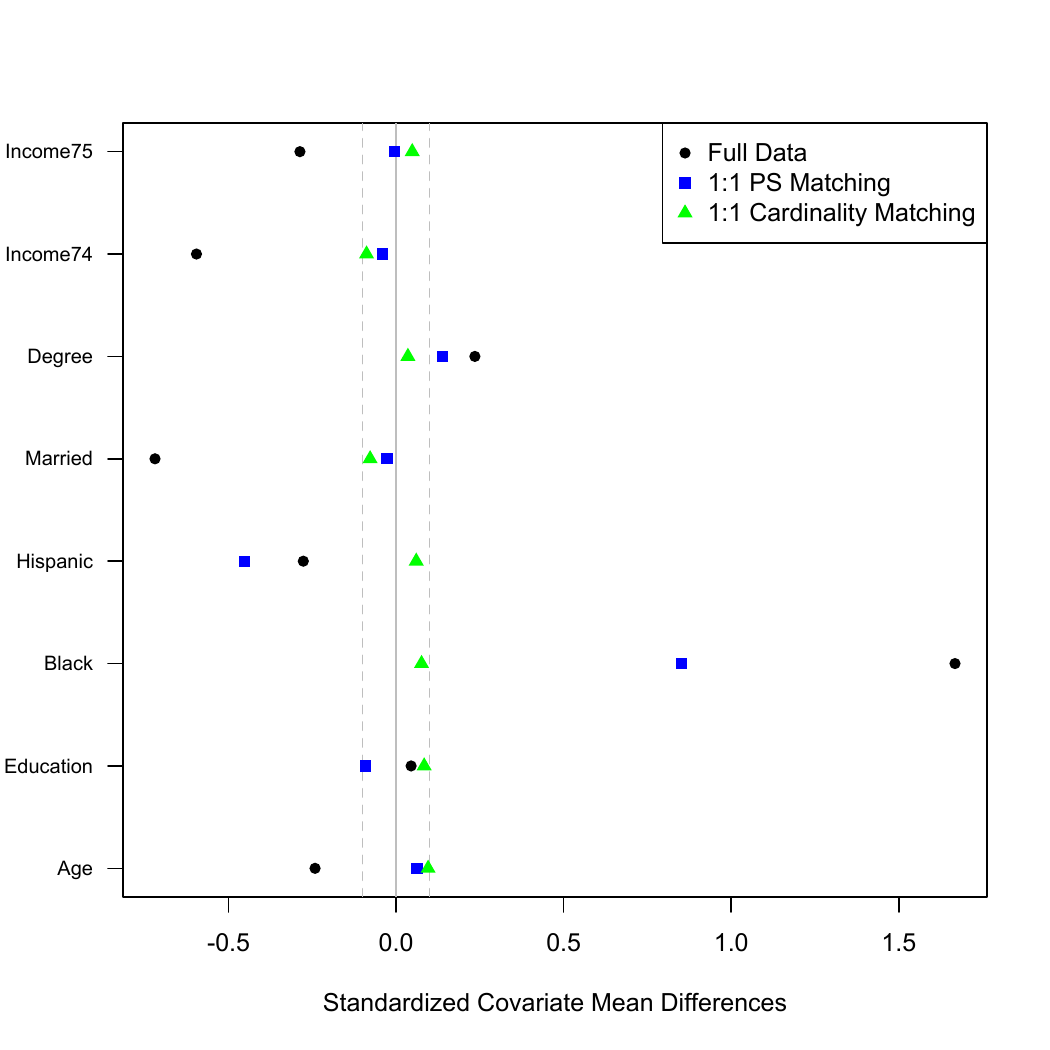}
  \caption{Love plot for the full Lalonde dataset, the 1:1 propensity score matched dataset, and the 1:1 cardinality matching dataset.}
  \label{fig:lalondeLovePlot}
\end{subfigure} \hfill
\begin{subfigure}[t]{0.45\textwidth}
\centering
  \includegraphics[width=1.1\textwidth]{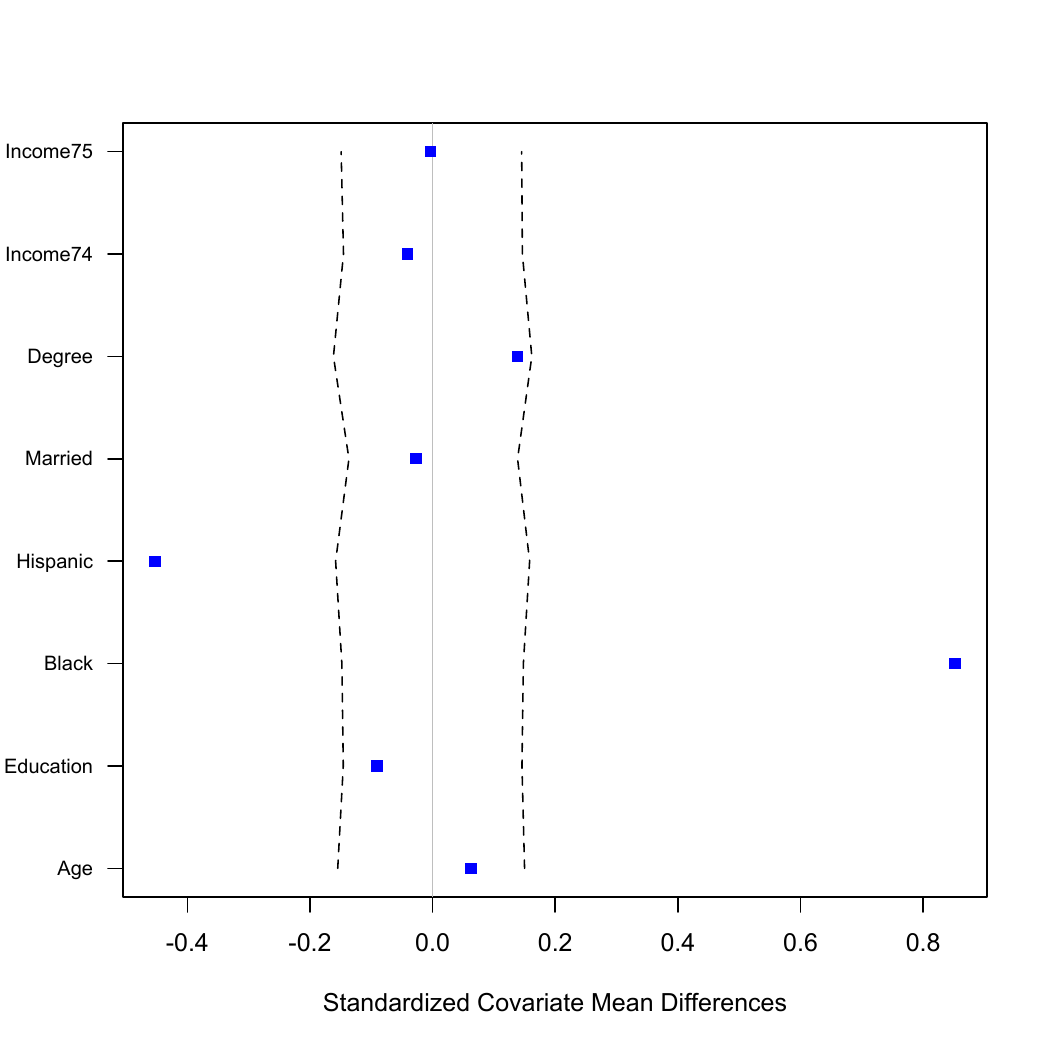}
  \caption{Love plot for the 1:1 propensity score matched dataset with 7.5\% and 92.5\% complete randomization quantiles denoted by dashed lines.}
  \label{fig:lalondePermPlot}
\end{subfigure}
\caption{Assessing balance of matched datasets using Love plots.}
\end{figure}

\begin{figure}
  \centering
  \includegraphics[scale=0.5]{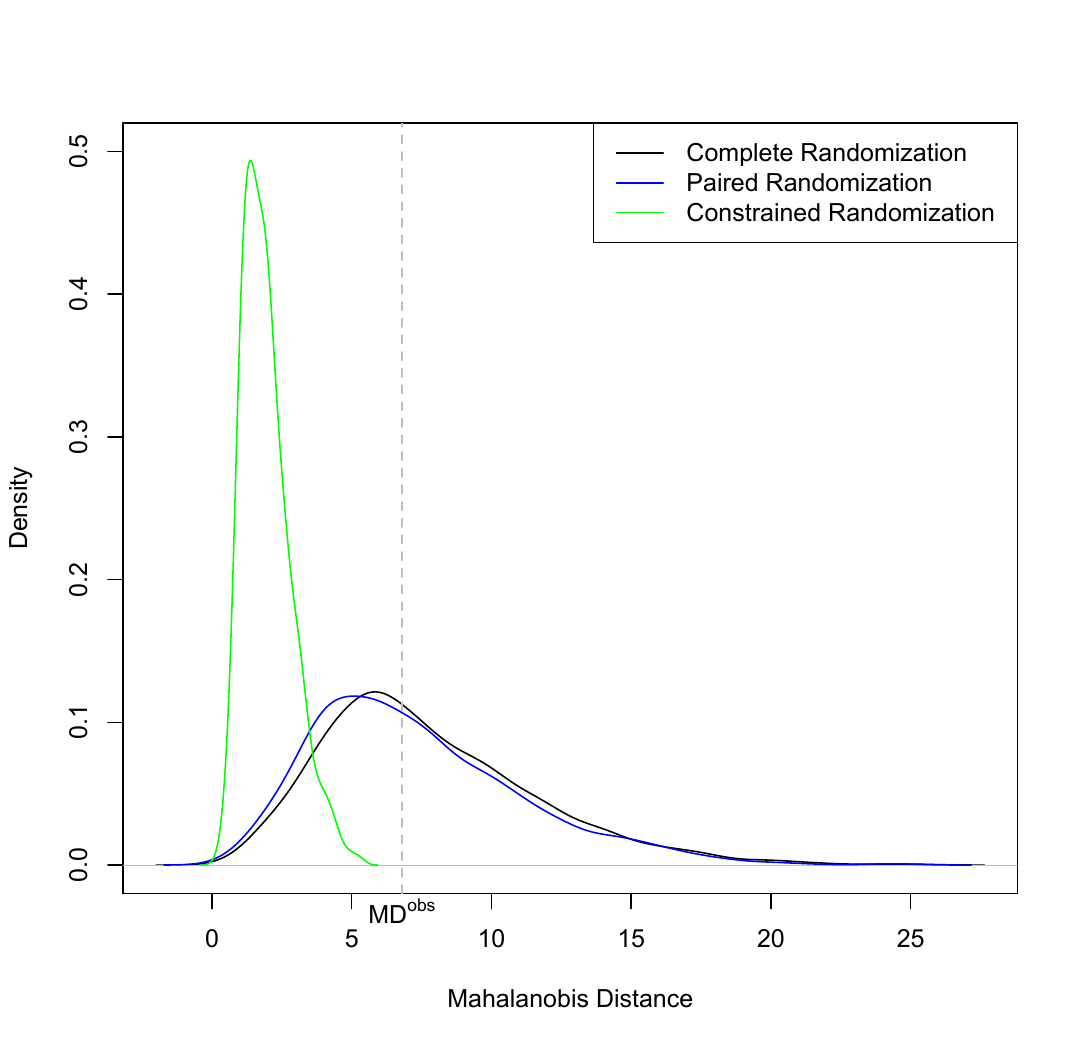}
  \caption{Distribution of the Mahalanobis distance across 1,000 complete randomizations, paired randomizations, and constrained randomizations for the 1:1 cardinality matched dataset. The observed Mahalanobis distance is denoted by a dashed vertical line.}
  \label{fig:lalondeMDPlot}
\end{figure}

Figure \ref{fig:lalondeLovePlot} shows the standardized covariate mean differences for these three datasets. There are large imbalances in the full dataset, which motivates matching methods. The matched datasets both exhibit degrees of covariate balance. First we ask: Is it reasonable to assume the propensity score matched dataset approximates a completely randomized experiment? To assess this, we run our randomization test under Complete Randomization by permuting $\mathbf{W}^{obs}$ 1,000 times and computing the standardized covariate mean differences for each permutation. Figure \ref{fig:lalondePermPlot} shows the 7.5\% and 92.5\% quantiles of each difference across these permutations; this corresponds to setting $\alpha = 0.15$ within our randomization test, using the standardized covariate mean differences as the test statistic. Any difference outside of these quantiles is considered surprisingly large under Complete Randomization. Because there are some differences outside of these quantiles, we conclude that Complete Randomization is not plausible for this dataset. Running our test using the Mahalanobis distance as a test statistic gave the same conclusion: The $p$-value was less than 0.001.

Now we ask: Is it reasonable to assume the cardinality matched dataset approximates a completely randomized experiment, or even another experimental design? Figure \ref{fig:lalondeMDPlot} provides a visual of our test using the Mahalanobis distance as a test statistic. We display the distribution of the Mahalanobis distance across 1,000 draws from Complete Randomization in (\ref{eqn:completeRandomization}), 1,000 draws from Paired Randomization in (\ref{eqn:blockRandomization}), and 1,000 draws from Constrained Randomization in (\ref{eqn:constrainedRandomization}). We see that the observed Mahalanobis distance is reasonably within the first two distributions, giving credence to assuming Complete Randomization or Paired Randomization (the corresponding $p$-values are 0.53 and 0.49, respectively). However, the observed Mahalanobis distance is outside of the Constrained Randomization distribution, suggesting that this experimental design is not plausible. At first this may be surprising: Constrained Randomization in (\ref{eqn:constrainedRandomization}) incorporates the constraint that all standardized covariate mean differences are below 0.1, and cardinality matching fulfills this constraint, so why does this design not appear plausible? There are two reasons. First, even though cardinality matching fulfilled this balance constraint, it did so just barely, as seen in Figure \ref{fig:lalondeLovePlot}. Thus, the matched dataset is unusually imbalanced according to this design. Second, the Mahalanobis distance accounts for the covariance among covariates; thus, the \textit{joint} imbalance in Figure \ref{fig:lalondeLovePlot} is considered unusual according to this design.

The above example demonstrates how our randomization test can make covariate-by-covariate or omnibus assessments of balance. In particular, we recommend using an omnibus assessment via the Mahalanobis distance, because it accounts for the joint relationship among covariates when assessing balance. Furthermore, using the Mahalanobis distance allows for a graphical display that places different designs on the same univariate scale, such that researchers can ascertain which design is most appropriate for a particular dataset. Finally, this example demonstrates that even when covariate balance constraints hold for a matched dataset, it still may not be appropriate to assume a balance-constrained design.

\section{The Analysis Stage: After Assuming an Experimental Design for Matched Data} \label{s:analysisStage}

Once a particular design is assumed for a matched dataset, causal analyses become relatively straightforward, to the extent that they are straightforward for an experiment that uses that design. There are Fisherian (also known as randomization-based), Neymanian, and Bayesian modes of inference for analyzing such experiments.

Randomization-based inference focuses on testing sharp hypotheses, such as $Y_i(1) = Y_i(0)$ for all $i=1,\dots,N$. Under a sharp hypothesis, the potential outcomes for any treatment assignment are known. Researchers can also invert sharp hypotheses to obtain point estimates and uncertainty intervals. One possible sharp hypothesis is that the treatment effect is additive, i.e., that $Y_i(1) = Y_i(0) + \tau$ for some $\tau \in \mathbb{R}$ for all $i=1,\dots,N$. Then, a randomization-based uncertainty interval is the set of $\tau$ such that one fails to reject this sharp hypothesis, and a randomization-based point estimate is the $\tau$ yielding the highest $p$-value \citep{hodges1963estimates,rosenbaum2002observational}. To test such hypotheses, one must specify an assignment mechanism (and methods from Section \ref{s:randomizationTest} can be used to specify this) and a test statistic (some kind of treatment effect estimator). The choice of assignment mechanism can be viewed as a design-based decision, and the choice of test statistic can be viewed as a model-based decision. See \cite{rosenbaum2002observational} and \cite{imbens2015causal} for a general discussion of randomization-based inference. A limitation of this mode of inference is that a sharp null hypothesis must be specified, which often requires assuming away heterogeneous treatment effects, as in the aforementioned additive sharp hypothesis. See \citet[Section 4]{samii2012equivalencies} for further discussion on this limitation and \cite{caughey2016beyond} for an approach that incorporates treatment effect heterogeneity for this mode of inference. Despite this limitation, because randomization-based inference can flexibly incorporate different design-based and model-based decisions, we will focus primarily on this mode of inference in Sections \ref{s:simulations} and \ref{s:realDataAnalysis}.

Meanwhile, the Neymanian mode of inference involves asymptotic approximations for treatment effect estimators under certain experimental designs. There are well-known results on Neymanian inference for many experimental designs. For example, \cite{miratrix2013adjusting} discusses Neymanian inference for blocked experiments, and \cite{imai2008variance} does the same for paired experiments. See \cite{pashley2017insights} for a discussion of variance estimation for these two designs as well as hybrid designs that involve blocks and pairs. Neymanian inference has also been established for factorial designs \citep{dasgupta2015causal}, rerandomized experiments \citep{li2018asymptotic}, and their combination \citep{li2020rerandomizationFactorial}. See \cite{ding2017paradox} for a comparison of randomization-based and Neymanian modes of inference for blocked, matched-pair, and factorial designs, as well as \citet[Section 1]{fogarty2020studentized} for a historical, agnostic view on these modes of inference.

The seminal paper by \cite{li2018asymptotic} is particularly relevant to this work. \cite{li2018asymptotic} derived the asymptotic distribution of the mean-difference estimator for rerandomized experiments where the Mahalanobis distance is constrained to be below some threshold (an experimental design first proposed by \cite{morgan2012rerandomization}). This rerandomization scheme is similar to (but not the same as) Constrained Randomization (defined in (\ref{eqn:constrainedRandomization})). A promising line for future work is extending the results of \cite{li2018asymptotic} to designs like Constrained Randomization. For example, \cite{wang2019large} established several large-sample properties of matching methods that ensure covariate balance constraints hold by design. Continuing such developments would provide a useful way to conduct Neymanian inference for matched datasets using nuanced designs like Constrained Randomization.

Finally, the Bayesian mode of inference for estimating causal effects was first formalized in \cite{rubin1978bayesian}. Under this mode of inference, $(\mathbf{X}, \mathbf{W}, \mathbf{Y})$ are treated as unknown, and models for these quantities must be specified. After assuming $\mathbf{W} \sim P^*(\mathbf{W} | \mathbf{X})$, the remaining work for a Bayesian analysis is to specify statistical models for the covariates and outcomes. This mode of inference is particularly useful for incorporating uncertainty in many complex data scenarios, such as noncompliance \citep{frangakis2002clustered}, missing data \citep{rubin1996multiple}, their combination \citep{barnard2002school}, and multi-site trials \citep{dehejia2003there}; however, these complications are outside the scope of this work. See \cite{imbens2004nonparametric} and \cite{heckman2014treatment} for discussions of Bayesian inference for randomized experiments.

In what follows, we demonstrate how our randomization test can assess the plausibility of experimental designs for matched datasets, first in simulation (Section \ref{s:simulations}) and then in a real data analysis (Section \ref{s:realDataAnalysis}). In these sections we also consider standard Neymanian approaches for analyzing matched datasets assuming complete randomization or paired randomization, as well as a randomization-based approach for analyzing matched datasets assuming designs that leverage covariate balance, such as constrained randomization.

\section{Simulations} \label{s:simulations}

In this section, we will simulate datasets that exhibit covariate imbalance, thereby motivating matching methods. We will demonstrate how our test in Section \ref{s:randomizationTest} can be used to assess complete randomization, paired randomization, and constrained randomization designs for matched datasets. Furthermore, we will demonstrate how assuming constrained randomization for matched datasets can improve the precision of causal analyses when researchers match on covariates that are related to the outcomes. However, we will also see that assuming a precise experimental design can harm inferential results if there are still substantial biases due to covariate imbalances that remain after matching.

\subsection{Simulation Setup} \label{ss:simulationSetup}

We follow the simulation setup of \cite{austin2009some} and \cite{resa2016evaluation}, which has been used to evaluate different matching methods. Consider a dataset with $N_T = 250$ treated subjects and $N_C = 500$ control subjects. Each subject has four Normally distributed covariates and four Bernoulli distributed covariates. These eight covariates are generated such that the true standardized mean difference is 0.2 for half of the covariates and 0.5 for the other half. Furthermore, there are three outcomes: The first outcome is a linear function of the covariates, and the other two are nonlinear functions, where the third outcome is a more complex function than the second.\footnote{When each outcome was regressed on the covariates, $R^2 = 0.9$, $R^2 = 0.5$, and $R^2 = 0.25$ for the first, second, and third outcomes, respectively, on average across the 1,000 replicated datasets.} For each outcome, there is an additive treatment effect of one, which is the causal estimand in this simulation. Details about this data generating process are in Appendix \ref{s:appendixB}. By repeating this data-generating process, we produced 1,000 datasets with severe covariate imbalances.

\subsection{Design and Analysis Results for Complete Randomization and Paired Randomization} \label{ss:standardStatisticalInferenceSimulation}

In their simulation study, \cite{resa2016evaluation} compared nearest-neighbor propensity score matching, optimal subset matching \citep{rosenbaum2012optimal}, and cardinality matching \citep{zubizarreta2014matching}. Cardinality matching is similar to optimal subset matching in that it may discard some treated subjects in the name of achieving better balance; however, it differs from optimal subset matching in that it ensures group-level balance directly. \cite{resa2016evaluation} found that cardinality matching performs better than nearest-neighbor and optimal subset matching in terms of covariate balance, sample size, bias, and root mean squared error (RMSE). Thus, we focus on cardinality matching, and defer to \cite{resa2016evaluation} and other simulation studies (e.g., \cite{austin2009some} and \cite{austin2014comparison}) for a comparison of other methods.

When implementing cardinality matching, we focus on creating the largest pair-matched dataset such that the absolute standardized covariate mean differences are less than some threshold $a$; we consider $a = 0.1$ and $a = 0.01$. A more stringent threshold results in better balance but possibly a smaller sample size.\footnote{When $a = 0.1$, sample sizes ranged from 438 to 500, and when $a = 0.01$, they ranged from 386 to 494. A sample size less than 500 means some treated subjects were discarded in the name of achieving better balance. Discarding treated subjects changes the causal estimand, but this isn't problematic when the treatment effect is homogeneous (as is the case here).} Doing this produced 1,000 matched datasets with threshold $a = 0.1$ and 1,000 datasets with threshold $a = 0.01$. Table \ref{tab:biasMSECoverage} shows the bias, variance, and RMSE of the mean-difference estimator across these matched datasets. Table \ref{tab:biasMSECoverage} also shows the coverage of what we call ``complete randomization 95\% confidence intervals'' and ``paired randomization 95\% confidence intervals,'' which are computed as
\begin{align}
  \underbrace{\hat{\tau} \pm 1.96 \sqrt{ \frac{\hat{\sigma}^2_T}{N_T} + \frac{\hat{\sigma}^2_C}{N_C} }}_{\text{complete randomization}} \hspace{0.5 in} \text{ and } \hspace{0.5 in}
  \underbrace{\hat{\tau} \pm 1.96 \sqrt{ \frac{\sum_{j=1}^J (\hat{\tau}_j - \hat{\tau})^2}{J(J-1)} }}_{\text{paired randomization}} \label{eqn:completePairedRandomizationCIs}
\end{align}
where $\hat{\tau}$ is the mean-difference estimator, $\hat{\tau}_j$ is the mean-difference estimator within pair $j = 1,\dots,J$, and $\hat{\sigma}^2_T$ and $\hat{\sigma}^2_C$ are the sample variances in treatment and control. These are the standard Neymanian confidence intervals for the average treatment effect in completely randomized and paired experiments \citep{imai2008variance,imbens2015causal}.

When $a = 0.1$, there is substantial bias, resulting in confidence intervals undercovering. As expected, the paired randomization confidence intervals are narrower than the complete randomization confidence intervals, resulting in even worse coverage. Thus, it can be harmful to condition on the pairs of a matched dataset. However, when $a = 0.01$, the bias is neglible and the confidence intervals tend to overcover. This overcoverage is most prominent for the first outcome, followed by the second, and finally by the third. This is ordered by how ``well-specified'' cardinality matching was, which only attempted to balance the raw covariates---which define the first outcome---and not the nonlinear functions that define the second and third outcomes (as detailed in Appendix \ref{s:appendixB}).

Table \ref{tab:biasMSECoverage} show the inferential results when we assume complete randomization or paired randomization for the matched datasets, but were these assumptions appropriate? Note that all of the matched datasets fulfill the common rule-of-thumb that standardized covariate mean differences be below 0.1, and thus it would be common practice to assume complete randomization or paired randomization for these datasets. Alternatively, it is also common to perform balance tests to assess these assumptions; we will consider three here:
\begin{enumerate}
  \item Use a $t$-test for each of the eight covariates, and then define the balance test $p$-value as the minimum of the eight resulting $p$-values.
  \item Use a Kolmogorov-Smirnov test for each of the eight covariates, and then define the balance test $p$-value as the minimum of the eight resulting $p$-values.
  \item Use our randomization test from Section \ref{ss:randomizationTest}, using the Mahalanobis distance (\ref{eqn:md}) as the test statistic.
\end{enumerate}
For the first two balance tests, we choose the minimum $p$-value among the covariate-specific $p$-values in order to be conservative when assessing randomized assignment \citep{diamond2013genetic,cattaneo2015randomization}. Meanwhile, our randomization test using the Mahalanobis distance acts as a global test for covariate balance, thereby providing a single $p$-value. Table \ref{tab:rejectionRate} shows the rejection rate of complete randomization and paired randomization for these balance tests, where we reject if the $p$-value is less than $\alpha = 0.15$ (as recommended in \cite{cattaneo2015randomization} and our Section \ref{ss:randomizationTest}). For the $a = 0.1$ matched datasets, the $t$-test never rejects complete randomization or paired randomization, where we used the paired $t$-test to assess paired randomization. Meanwhile, the Kolmogorov-Smirnov test and our test reject complete randomization nearly half the time. Note that, unlike the $t$-test, there is not a paired version of the Kolmogorov-Smirnov test, and thus we do not display a paired randomization $p$-value. On the other hand, our test can assess any experimental design, and our test rejects over 75\% of the time for paired randomization. Because inferential results are quite biased when $a = 0.1$, it is reassuring that our test and the Kolmogorov-Smirnov test frequently reject randomized assignment for these matched datasets. Furthermore, this demonstrates the advantages of assessing forms of covariate balance beyond marginal means---e.g., joint mean balance like the Mahalanobis distance or marginal distribution balance via the Kolmogorov-Smirnov test. At the same time, this demonstrates that our approach is not a panacea: Another way to view Table \ref{tab:rejectionRate} is that our test incorrectly failed to reject complete randomization nearly half the time and paired randomization nearly a quarter of the time, thereby giving us false confidence in biased analyses. Thus, this also demonstrates that even if our test fails to reject a particular design, there's no guarantee that the resulting inference using such a design will be valid or unbiased. Meanwhile, the $t$-test and our test never reject complete randomization or paired randomization for the $a = 0.01$ matched datasets, although the Kolmogorov-Smirnov test rejects complete randomization 3.2\% of the time. Considering that inferential results are relatively unbiased and conservative, this lack of rejection is reassuring.

\begin{table}
\centering
  \begin{tabular}{cccccc}
  \hline
    \textbf{Outcome} & \textbf{Bias} & \textbf{Variance} & \textbf{RMSE} & \begin{tabular}{@{}c@{}} \textbf{Coverage} \\ \textbf{(CR 95\% CIs)}\end{tabular} & \begin{tabular}{@{}c@{}} \textbf{Coverage} \\ \textbf{(PR 95\% CIs)}\end{tabular} \\
    \hline
  \rowcolor{Gray}
  \textit{First Outcome} & & & & & \\
  $a = 0.1$ & 0.92  & 0.04  & 0.94  & 73.5\%  & 51.3\%  \\
  $a = 0.01$ & 0.05  & 0.04  & 0.20  & 100\%  & 100\%  \\
  \rowcolor{Gray}
  \textit{Second Outcome} & & & & & \\
  $a = 0.1$ & 2.10  & 0.58 & 2.24  & 58.1\%  & 50.1\%  \\
  $a = 0.01$ & 0.42  & 0.54  & 0.84  & 99.7\%  & 99.1\%  \\
  \rowcolor{Gray}
  \textit{Third Outcome} & & & & & \\
  $a = 0.1$ & 0.79  & 1.56 & 1.48  & 92.1\%  & 90.5\%  \\
  $a = 0.01$ & 0.01  & 1.40  & 1.18 & 98.1\% & 97.5\%  \\
  \hline
  \end{tabular}
  \caption{Properties of the mean-difference estimator after cardinality matching, where the standardized covariate mean differences are constrained to be less than some threshold $a$. Complete randomization (CR) and paired randomization (PR) intervals are defined in (\ref{eqn:completePairedRandomizationCIs}).}
  \label{tab:biasMSECoverage}
\end{table}

\begin{table}
\centering
  \begin{tabular}{c|cc}
  \hline
    & \textbf{CR $p$-value} & \textbf{PR $p$-value} \\
    \textbf{Dataset/Test} & \textbf{Rejection Rate} & \textbf{Rejection Rate} \\
    \hline
    \rowcolor{Gray}
    $a = 0.1$ & & \\
    $t$-test & 0.0\% & 0.0\% \\
    Kolmogorov-Smirnov test & 48.7\% & NA \\
    Randomization test & 45.2\% & 76.6\% \\
    \rowcolor{Gray}
    $a = 0.01$ & & \\
    $t$-test & 0.0\% & 0.0\% \\
    Kolmogorov-Smirnov test & 3.2\% & NA \\
    Randomization test & 0.0\% & 0.0\% \\
    \hline
  \end{tabular}
  \caption{Rejection rate for three balance tests---where we reject if the $p$-value is less than $\alpha = 0.15$---for the $a = 0.1$ and $a = 0.01$ cardinality matched datasets. By using a two-sample or paired $t$-test, we can assess complete randomization (CR) or paired randomization (PR); however, because there is not a paired version of the Kolmogorov-Smirnov test, it can only assess complete randomization. To be conservative when running the $t$-test and Kolmogorov-Smirnov test, we took the minimum of the eight covariate-specific $p$-values. Meanwhile, our test---by using the Mahalanobis distance---assesses global covariate balance with a single $p$-value, and also can assess any experimental design.}
  \label{tab:rejectionRate}
\end{table}

\subsection{Design and Analysis Results for Constrained Paired Randomization} \label{ss:morePreciseCausalAnalysesSimulations}

In the previous section we did not find evidence against complete randomization or paired randomization for the $a = 0.01$ cardinality matched datasets, which were designed to exhibit a high level of balance across all covariates. Now we'll consider the consequences of analyzing these matched datasets assuming a variant of Constrained Randomization:
\begin{align}
    P(\mathbf{W} = \mathbf{w} | \mathbf{X}) &= \begin{cases}
        |\mathcal{A}|^{-1} &\mbox{ if } \mathbf{w} \in \mathcal{A} \\
        0 &\mbox{ otherwise.} \label{eqn:constrainedRandomizationSimulation}
    \end{cases}
\end{align}
where $\mathcal{A} \equiv \{\mathbf{w}: |\bar{\mathbf{x}}_T - \bar{\mathbf{x}}_C | < 0.05 \text{ and } \sum_{i=1}^N w_i = N_T\}$. This is a natural design to posit for cardinality matching, because we constrained $|\bar{\mathbf{x}}_T - \bar{\mathbf{x}}_C |$ by design. To assess if this design is plausible for the $a = 0.01$ cardinality matched datasets, we ran our test using the Mahalanobis distance as a test statistic; the randomization test $p$-values were consistently greater than 0.8, suggesting the plausibility of this experimental design.

To analyze the $a = 0.01$ cardinality matched datasets assuming constrained randomization in (\ref{eqn:constrainedRandomizationSimulation}), we take a randomization-based approach and test the sharp hypothesis $H_0^{\tau}: Y_i(1) = Y_i(0) + \tau$ for $\tau \in \{-12.00, -11.99, \dots, 11.99, 12.00\}$. Define the 95\% randomization-based confidence interval as the set of $\tau$ such that we fail to reject $H_0^{\tau}$ at the $\alpha = 0.05$ level. We'll consider the mean-difference estimator and linear regression estimator as test statistics when inverting this hypothesis test.

The first three rows of Table \ref{tab:ciComparison} compare the randomization-based confidence intervals for the mean-difference estimator under constrained randomization to the complete randomization and paired randomization confidence intervals from the previous section. The coverage of the constrained randomization confidence intervals is closer to the nominal level, and they are substantially narrower: They are on average 45\% the width for the first outcome, 66\% the width for the second outcome, and 88\% the width for the third outcome, as compared to the complete randomization confidence intervals. However, the constrained randomization confidence intervals undercover for the second outcome. This is likely because of nonneglible bias, as seen in Table \ref{tab:biasMSECoverage}. Thus, in some sense it was lucky that the complete randomization and paired randomization confidence intervals overcovered for the second outcome despite this bias. This echoes the observation made in Table \ref{tab:biasMSECoverage}, where paired randomization undercovered even more so than complete randomization for the $a = 0.1$ cardinality matched datasets. Thus, it may be harmful to assume a precise experimental design or even any design if there are substantial biases that remain after matching. Indeed, an argument can be made to conservatively assume a less precise design---e.g., complete randomization---in the hope that the resulting wider confidence intervals will be closer to the nominal level if there are biases that remain after matching.

\newcolumntype{g}{>{\columncolor{Gray}}c}
\begin{table}
\centering
\resizebox{\columnwidth}{!}{%
  \begin{tabular}{cggccgg}
  \hline
  & \multicolumn{2}{g}{ \textit{First Outcome} } & \multicolumn{2}{c}{ \textit{Second Outcome} } & \multicolumn{2}{g}{ \textit{Third Outcome} } \\
    \textbf{CI Method} & \begin{tabular}{@{}c@{}} \textbf{Average} \\ \textbf{CI Length} \end{tabular} & \textbf{Coverage} & \begin{tabular}{@{}c@{}} \textbf{Average} \\ \textbf{CI Length} \end{tabular} & \textbf{Coverage} & \begin{tabular}{@{}c@{}} \textbf{Average} \\ \textbf{CI Length} \end{tabular} & \textbf{Coverage} \\
    \hline
    \textit{Mean-Difference Analysis} & & & & & & \\
    Complete Rand. & 2.14  & 100\%  & 4.55  & 99.7\% & 5.09 & 98.1\%  \\
    Paired Rand. & 1.93  & 100\%  & 4.21 & 99.1\%  & 4.95 & 97.5\%  \\
    Constrained Rand. & 0.96 & 97.9\%  & 2.94 & 91.5\%  & 4.46 & 95.5\% \\
    \hline
    \textit{Linear Regression Analysis} & & & & & &  \\
    Complete Rand. & 0.74  &  95.3\% &  2.71 &  90.8\% &  4.49 & 95.9\% \\
    Paired Rand. & 0.74 & 95.2\% & 2.65 & 88.9\% & 4.46 & 95.6\% \\
    Constrained Rand. & 0.73 & 94.3\%  & 2.71  & 90.2\%   & 4.40  & 95.4\%   \\
    \hline
  \end{tabular}
  }
  \caption{Average length and coverage of complete randomization, paired randomization, and constrained randomization confidence intervals for the mean-difference estimator and linear regression estimator for the $a = 0.01$ cardinality matched datasets.}
  \label{tab:ciComparison}
\end{table}

Meanwhile, the last three rows of Table \ref{tab:ciComparison} show results for the treatment effect estimator from a linear regression of the outcomes on $\mathbf{X}$ and $\mathbf{W}$. The complete randomization confidence intervals were computed using the standard error of the regression coefficient for $\mathbf{W}$. The paired randomization confidence intervals were computed using the variance estimator of \cite{fogarty2018regression}, who provides recent results on regression adjustment for paired experiments.\footnote{We used the variance estimator $S^2_{R1}$ in \cite{fogarty2018regression}, which utilizes pairwise-differences of functions of the covariates among matched pairs. We specified this function as the raw covariates, $(\mathbf{x}_1,\dots,\mathbf{x}_8)$, to make the paired randomization analysis comparable to the complete randomization analysis. \cite{fogarty2018regression} also discusses utilizing pairwise-\textit{averages} of functions of the covariates among matched pairs, which we do not consider here.} In this case, the results across experimental designs are nearly identical. When the covariates are linearly related with the outcome---as is the case for the first outcome---coverage for linear regression is close to the nominal level. However, there is undercoverage for the second outcome, which is a nonlinear function of the covariates. This provides two findings. First, linear regression after matching is not guaranteed to exhibit the correct coverage. Second, as discussed in Section \ref{s:analysisStage}, the assignment mechanism in our randomization-based approach can be viewed as a design-based choice, while the treatment effect estimator is a model-based choice. Table \ref{tab:ciComparison} suggests that there may be an equivalence among certain design-based and model-based choices, which echoes recent equivalences found between rerandomization designs and regression adjustment \citep{li2020rerandomization}.

In summary, our test can assess standard experimental designs like complete randomization and paired randomization, as well as nuanced designs like constrained randomization. When matched datasets exhibit high levels of covariate balance, complete randomization and paired randomization analyses can be conservative, and designs that account for strong covariate balance can provide more precise causal analyses. However, using a precise experimental design like paired randomization or constrained randomization should be proceeded with caution, because it can harm inferential results if there are still substantial biases due to covariate imbalances that were not accounted for in the matching stage. Nonetheless, the additional precision from precise experimental designs can be substantial if researchers match on relevant covariates. In particular, subject-matter expertise often guides researchers towards balancing certain covariates that are deemed relevant a priori. In Section \ref{s:realDataAnalysis}, we revisit a causal analysis conducted by \cite{keele2017black}, who used matching to target balancing certain covariates. We will demonstrate how our randomization test can assess the type of experimental design their matched dataset approximates, and how a constrained randomization design can improve precision for this application.

\section{Revisiting a Causal Analysis of the Effects of Candidates' Race on Black Voter Turnout} \label{s:realDataAnalysis}

An ongoing problem in political science is determining how minority candidates affect minority voter turnout in American elections. Many works have found a positive relationship between minority candidate participation in elections and minority voter turnout; to explain this phenomenon, these works argue that minorities feel empowered when they witness a minority candidate run for political office, thereby increasing voter turnout \citep{browning1984protest,bobo1990race,leighley2001strength,barreto2004mobilizing}. However, these findings have primarily been correlational instead of causal.

Recently, \cite{keele2017black} addressed this research question using matching to conduct a causal analysis assessing if having at least one African American candidate in Louisiana mayoral elections affected black voter turnout.\footnote{Following the practice of the United States Census Bureau and \cite{keele2017black}, we use ``African American'' and ``black'' interchangeably.} We will apply our randomization test to the \cite{keele2017black} matched dataset to assess if this dataset approximates a particular experimental design, use our randomization-based inferential approach to analyze the matched dataset, and compare our approach to the more standard approach used in \cite{keele2017black}. Revisiting this causal analysis is particularly suitable for assessing our approach for two reasons. First, \cite{keele2017black} used cardinality matching, which is the method we focused on in Section \ref{s:simulations}. Second, and more importantly, \cite{keele2017black} used subject-matter expertise to target achieving high levels of balance on certain covariates, and our approach can condition on these high levels of covariate balance to provide a precise causal analysis. First, we describe the full data and matched data in \cite{keele2017black}. Then, we compare our approach to a standard approach for analyzing these data.

\subsection{Description of the Full Dataset and Matched Dataset}

The data include 1,006 mayoral elections in Louisiana from 1988-2011. Data is at the municipality level, where each election was held. Covariates include each municipality's median income, number of residents, percentage of residents that are African American and of voting age, percentage of residents with a college degree, percentage of residents with a high school degree, percentage of residents that are unemployed, percentage of residents that are below the poverty line, and whether or not it had a home rule charter.\footnote{These covariates are based on 1990 census data and thus are pre-treatment measurements for most of the electoral data.} The treatment is whether or not at least one candidate in the election was African American. The outcome is black voter turnout (measured in percentage points), and interest is in the ATE on this outcome. \cite{keele2017black} created this dataset using three data sources maintained by the state of Louisiana, and further details can be found therein.

\cite{keele2017black} focused their efforts on these data because all mayoral elections in Louisiana can turn into a runoff election. In Louisiana mayoral elections, a ``general election'' is held at first, where any number of candidates may run. If no candidate receives the majority of votes, the two candidates with the most votes advance to a ``runoff election.'' \cite{keele2017black} analyzed general elections as well as runoff elections. For ease of exposition, we focus on general elections, because \cite{keele2017black} were able to achieve a larger matched sample and higher level of covariate balance for these data.

The full dataset exhibits large covariate imbalances (see Table 1 of \cite{keele2017black}). In particular, treated municipalities have substantially higher proportions of African American residents. \cite{keele2017black} posited that this covariate was particularly relevant to black voter turnout, and so they used cardinality matching to ensure strong balance on this covariate, as well as balance on the other covariates. Furthermore, they ensured near-exact balance \citep[Chapter 9]{rosenbaum2010design} on election year by creating pairs of treatment and control elections that occurred during the same year or one year after each other. This is an example of the common practice of using subject-matter expertise to prioritize certain forms of covariate balance when designing a matched dataset \citep{ramsahai2011extending,zubizarreta2012using,pimentel2015large,keele2020comparing}. The resulting matched dataset consisted of 197 pairs of elections that exhibited (1) near-exact balance on election year within each pair, (2) high balance on percentage of African American residents across pairs, and (3) balance on all other covariates across pairs. Table \ref{tab:balanceGeneralElections} shows the standardized covariate mean differences for each covariate in this dataset; all of the differences are below 0.1.\footnote{The municipal population covariate had a standardized mean difference of 0.097, but due to rounding it is displayed as 0.1 in Table \ref{tab:balanceGeneralElections}.}

\subsection{Which Experimental Design Does This Matched Dataset Approximate?}

\cite{keele2017black} argued that their matched dataset approximated a paired experiment by finding non-significant Kolmogorov-Smirnov tests for each covariate. This is a standard diagnostic in the matching literature, but it is not a valid test for a specific experimental design. To provide a valid test, we ran our randomization test for complete randomization and two paired designs:
\begin{align}
    P(\mathbf{W} = \mathbf{w} | \mathbf{X}) &= \begin{cases}
      {394 \choose 197}^{-1} &\mbox{ if } \sum_{i=1}^{394} w_i = 197 \\
      0 &\mbox{ otherwise.}
    \end{cases} \text{(Complete Randomization)} \label{eqn:completeRandomizationRealData} \\
    P(\mathbf{W} = \mathbf{w} | \mathbf{X}) &=  \begin{cases}
   \frac{1}{2^{197}} &\mbox{ if } \sum_{i \in \mathcal{B}_j} w_i = 1 \hspace{0.05 in} \forall j=1,\dots,197 \\
  0 &\mbox{ otherwise.}
  \end{cases} \text{(Paired Randomization)} \label{eqn:pairedRandomizationRealData} \\
  P(\mathbf{W} = \mathbf{w} | \mathbf{X}) &= \begin{cases}
        |\mathcal{A}^{(cp)}|^{-1} &\mbox{ if } \mathbf{w} \in \mathcal{A}^{(cp)} \\
        0 &\mbox{ otherwise.}
    \end{cases} \hspace{0.1 in}\text{(Constrained Paired Randomization)} \label{eqn:constrainedPairedRandomizationRealData}
\end{align}
where $\mathcal{A}^{(cp)} \equiv \{\mathbf{w}: \sum_{i \in \mathcal{B}_j} w_i = 1 \forall j =1,\dots,197, \hspace{0.1 in} |\bar{\mathbf{x}}_T - \bar{\mathbf{x}}_C | < 0.15, \text{ and } |\bar{\mathbf{x}}_T - \bar{\mathbf{x}}_C |^{(AA)} < 0.01 \}$ is the set of \textit{constrained paired randomizations}, where $|\bar{\mathbf{x}}_T - \bar{\mathbf{x}}_C |^{(AA)}$ denotes the standardized mean difference for the African American (\%) covariate. Complete Randomization in (\ref{eqn:completeRandomizationRealData}) does not account for covariate balance; Paired Randomization in (\ref{eqn:pairedRandomizationRealData}) recognizes that \cite{keele2017black} constructed pairs of elections with near-exact balance on election year; and Constrained Paired Randomization in (\ref{eqn:constrainedPairedRandomizationRealData}) additionally recognizes that certain levels of balance were achieved across pairs, particularly for the African American (\%) covariate. To run our randomization test, we generated 10,000 random draws from the three above assignment mechanisms and computed the standardized covariate mean differences for each draw; then, the randomization test $p$-value in (\ref{eqn:pvalue}) is defined as the proportion of absolute standardized covariate mean differences that are greater than the observed one. Table \ref{tab:balanceGeneralElections} shows the resulting $p$-values for each covariate and design. Our test indicates that all of the covariates are well-balanced according to these designs.

\begin{table}
\centering
  \begin{tabular}{ccccc}
  \hline
    \textbf{Covariate} & $\bar{\mathbf{x}}_T - \bar{\mathbf{x}}_C$ & \textbf{CR $p$-value} &  \textbf{PR $p$-value} & \textbf{Constr. PR $p$-value} \\
    \hline
    \rowcolor{Gray}
    Municipal Population & 0.10 & 0.34 & 0.30 & 0.21 \\
    African American (\%) & 0.00 & 0.99 & 0.99 & 0.92 \\
    \rowcolor{Gray}
    College degree (\%) & 0.00 & 0.96 & 0.94 & 0.93 \\
    High school degree  (\%) & -0.01 & 0.96 & 0.94 & 0.93 \\
    \rowcolor{Gray}
    Unemployed (\%) & 0.02 & 0.83 & 0.76 & 0.73 \\
    Below poverty line (\%) & 0.03 & 0.79 & 0.65 & 0.57 \\
    \rowcolor{Gray}
    Median income & -0.02 & 0.88 & 0.78 & 0.76 \\
    Home Rule & 0.05 & 0.55 & 0.38 & 0.36 \\
    \hline
  \end{tabular}
  \caption{Standardized covariate mean differences and randomization test $p$-values for Complete Randomization (CR), Paired Randomization (PR), and Constrained PR for the \cite{keele2017black} matched dataset.}
  \label{tab:balanceGeneralElections}
\end{table}

\begin{figure}
\centering
  \includegraphics[scale=0.55]{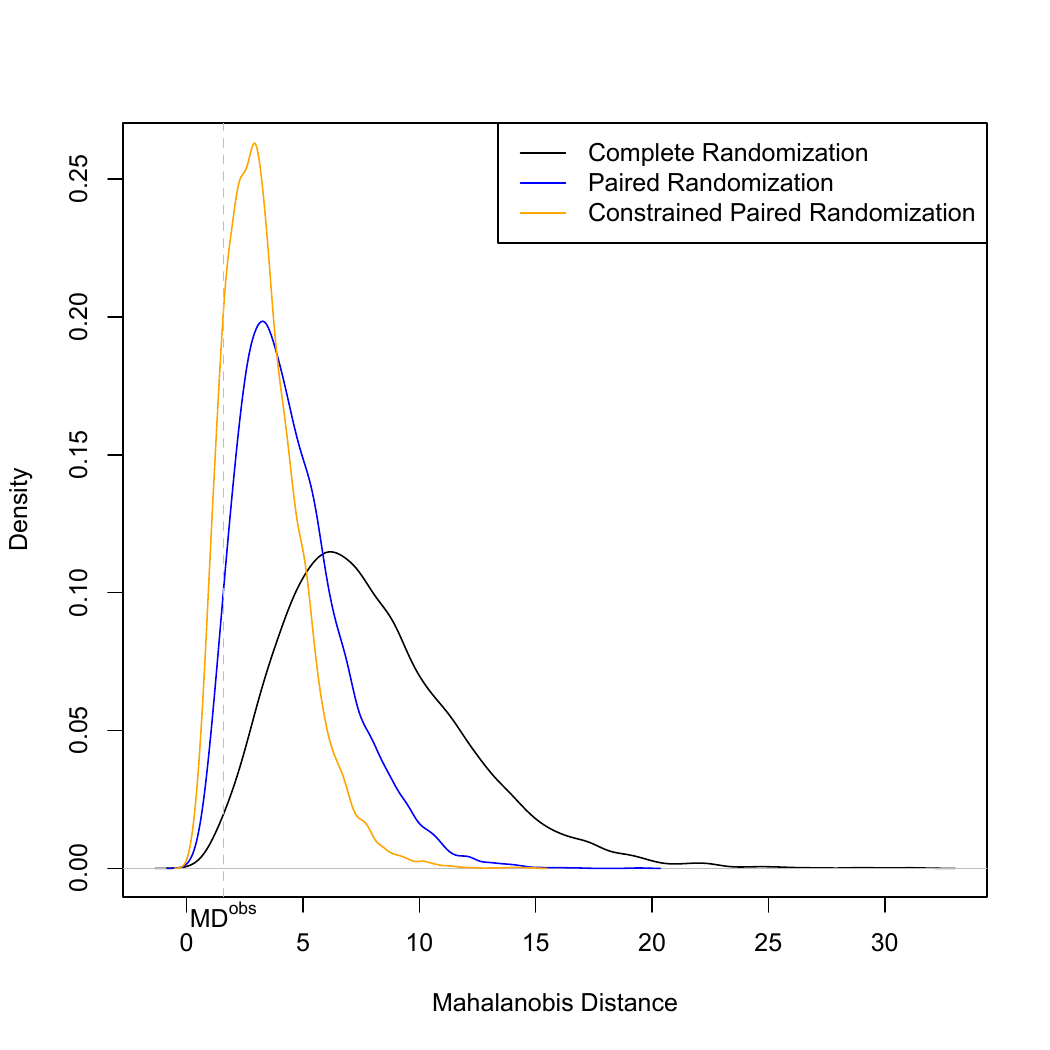}
  \caption{Distribution of the Mahalanobis distance for different designs. We generated 10,000 random draws from Complete Randomization in (\ref{eqn:completeRandomization}), Paired Randomization in (\ref{eqn:pairedRandomizationRealData}), and Constrained Paired Randomization in (\ref{eqn:constrainedPairedRandomizationRealData}); computed the Mahalanobis distance of each draw; and plotted the corresponding distributions and the observed Mahalanobis distance in the matched dataset. The corresponding $p$-values are 0.99 for Complete Randomization, 0.95 for Paired Randomization, and 0.87 for Constrained Paired Randomization.}
  \label{fig:keeleMDPlot}
\end{figure}

To determine which of these designs is most appropriate for this matched dataset, we computed the Mahalanobis distance across 10,000 draws from Complete Randomization, Paired Randomization, and Constrained Paired Randomization. Figure \ref{fig:keeleMDPlot} shows the resulting distribution of Mahalanobis distances under these three designs, as well as the observed Mahalanobis distance for the matched dataset. We can see that the matched dataset is unusually well-balanced compared to what we would expect from Complete Randomization and Paired Randomization, whereas the observed balance is near the mode of what we would expect from Constrained Paired Randomization, indicating that it may be the most appropriate design for this matched dataset. The corresponding $p$-values are 0.99 for Complete Randomization, 0.95 for Paired Randomization, and 0.87 for Constrained Paired Randomization, again indicating the plausibility of these three designs.

\subsection{Causal Analyses Under Different Experimental Designs}

Now we will analyze the matched data under Complete Randomization, Paired Randomization, and Constrained Paired Randomization, all of which were found to be plausible in the previous section. As in Section \ref{s:simulations}, we will construct randomization-based confidence intervals for this matched dataset by inverting the hypothesis $H_0^{\tau}: Y_i(1) = Y_i(0) + \tau$ using the mean-difference estimator as the test statistic. Table \ref{tab:generalElectionAnalyses} shows our randomization-based confidence intervals assuming Complete Randomization, Paired Randomization, and Constrained Paired Randomization, as well as the confidence interval reported in \cite{keele2017black}. All of the confidence intervals suggest that the presence of at least one African American candidate in Louisiana mayoral elections significantly increases black voter turnout.

\begin{table}
  \centering
  \begin{tabular}{cc}
    \hline
    \textbf{Analysis} & \textbf{Confidence Interval} \\
    \hline
    Keele et al. & (0.74, 6.04) \\
    Complete Rand. & (0.09, 6.75)  \\
    Paired Rand. & (0.70 6.11)  \\
    Constrained Paired Rand. & (0.81 6.00)  \\
    \hline
    \end{tabular}
    \caption{Confidence intervals for the ATE after matching. Units are in percentage points.}
    \label{tab:generalElectionAnalyses}
\end{table}

The confidence interval in \cite{keele2017black} was constructed by inverting Wilcoxon's signed rank test, which is a common approach for analyzing matched-pair data \citep{rosenbaum2002observational}. This is why our randomization-based confidence interval assuming Paired Randomization is quite similar to the \cite{keele2017black} confidence interval. These analyses condition on the matched pairs---which is why they are substantially narrower than the complete randomization confidence interval---but they do not condition on the covariate balance across pairs. Meanwhile, Constrained Paired Randomization does, resulting in a slightly narrower confidence interval: It is approximately 78\% the width of the complete randomization confidence interval and 96\% the width of the paired randomization confidence interval. This is in line with the width reduction we saw via simulation in Section \ref{s:simulations} for the third outcome, which was only somewhat linearly related with the covariates in that simulation ($R^2 = 0.25$ on average across the 1,000 simulated datasets). In this application, when we regressed black voter turnout on all the covariates in the full dataset, $R^2 = 0.18$, and when we regressed black voter turnout on all the covariates but election year (which was almost exactly matched in each pair), $R^2 = 0.12$. The graphical diagnostic in Figure \ref{fig:keeleMDPlot} provides evidence in favor of using the narrower confidence interval assuming Constrained Paired Randomization for this dataset. Even though all of the confidence intervals in Table \ref{tab:generalElectionAnalyses} suggest the same conclusion for this application, they demonstrate how assuming different designs within an observational study can lead to different levels of precision.

However, an important contribution of \cite{keele2017black} was conducting a sensitivity analysis to assess how sensitive their results were to hidden biases; they were able to conduct such an analysis using tools that assume a paired design (e.g., \citealt[Chapter 4]{rosenbaum2002observational}). A promising line of future work is extending these tools to more complex designs, like Constrained Paired Randomization.

\section{Discussion and Conclusion} \label{s:conclusion}

Covariate imbalance is one of the principal problems of causal inference in observational studies. To tackle this problem, matching algorithms can produce datasets with strong covariate balance. It is common to assume matched datasets approximate completely randomized or block randomized experiments if covariate balance diagnostics are met, even though these diagnostics do not formally assess whether treatment is effectively randomized. Instead, we propose a randomization test for assessing if a particular experimental design is plausible for a matched dataset. Our test is a generalization of other randomization tests for assessing covariate balance \citep{hansen2008covariate,cattaneo2015randomization,gagnon2019classification}, where we can test any experimental design, including designs with covariate balance constraints. In the analysis stage, we recommend a randomization-based approach, which can flexibly incorporate any assignment mechanism---a design-based decision---and any treatment effect estimator---a model-based decision.

Our test, like all balance tests, is limited in that failing to reject does not ``prove'' that treatment is effectively randomized or that assuming randomized assignment is guaranteed to yield valid inferential results. Nonetheless, our test serves as a helpful tool for detecting clear violations of random assignment that harm inferential results for matched data. In particular, we recommend using the Mahalanobis distance as a test statistic, because it accounts for the joint behavior of covariates while acting as a global test for balance. Meanwhile, we found that well-designed matched datasets that exhibit high levels of covariate balance tend to approximate balance-constrained designs like rerandomization. Analyzing these matched datasets as such can lead to precise causal analyses. However, assuming a precise experimental design for a matched dataset should be proceeded with caution, because it can harm inferential results if there are still imbalances in relevant covariates after matching.

To demonstrate how to use our balance diagnostics and inferential approach in practice, we revisited a causal analysis conducted in political science by \cite{keele2017black}. Researchers often have field-specific knowledge guiding them towards balancing particular covariates when matching, which was the case in this application. Using our approach, we pinpointed the experimental designs the resulting matched dataset may approximate. Furthermore, we improved the precision of this causal analysis by conditioning on the high levels of balance that researchers---using subject-matter expertise---ensured by design.

Because our framework combines design-based and model-based decisions, a promising line of future research is comparing different combinations of these decisions and assessing which combinations yield the best inference for a matched dataset. In particular, our simulations suggest that there may be an equivalence between certain designs when linear regression is used. This echoes recent findings that design-based matching methods and model-based machine learning methods often yield similar results \citep{keele2020comparing}. Furthermore, our approach can be applied to settings beyond matching. For example, assumptions of random assignment have been used in regression discontinuity designs \citep{li2015evaluating,cattaneo2015randomization,matteiMealli2016,branson2018local} and instrumental variable approaches \citep{brookhart2007preference,baiocchi2014instrumental,branson2020evaluating}. Our \texttt{randChecks} \texttt{R} package---used throughout this paper---can formally test effective random assignment of any binary indicator, such as binary treatments in regression discontinuity designs and binary instrumental variables.

% Acknowledgements should go at the end, before appendices and references

\section*{Acknowledgements}

We would like to thank Stephen Blyth, Luis Campos, Lucas Janson, Edward Kennedy, Luke Miratrix, Reagan Mozer, Nicole Pashley for insightful comments that led to substantial improvements in this work. We would also like to thank Luke Keele for providing the data used in our application. This research was supported by the National Science Foundation Graduate Research Fellowship Program under Grant No. 1144152. Any opinions, findings, and conclusions or recommendations expressed in this material are those of the authors and do not necessarily reflect the views of the National Science Foundation.


\newpage

\setcounter{section}{1}

\appendix
\section{Replication Code for the Lalonde Example} \label{s:appendixA}

Here we demonstrate how the tests and visuals presented in Section \ref{ss:graphicalDiagnostic} were created using our \texttt{randChecks} \texttt{R} package. This also serves as a demonstration as to how \texttt{randChecks} can be used to assess random assignment of a binary indicator, e.g., a treatment within a matched dataset.

The \texttt{randChecks} \texttt{R} package is available on \texttt{CRAN}; after installing the package, one can load the three datasets in Section \ref{ss:graphicalDiagnostic} using the following line of code:
\begin{verbatim}
  load("lalondeMatches.rda")
\end{verbatim}
This loads three \texttt{R} objects:
\begin{itemize}
  \item \texttt{lalonde}: The full Lalonde dataset of 614 subjects, 185 of whom are in treatment and 429 are in control.
  \item \texttt{lalonde.matched.ps}: The 1:1 propensity score matched dataset created using the \texttt{MatchIt} \texttt{R} package. This dataset contains 370 subjects, 185 of whom are in treatment and 185 are in control. Thus, compared to the full Lalonde dataset, this dataset contains the full treatment group but a subset of the control group.
  \item \texttt{lalonde.matched.card}: The 1:1 cardinality matched dataset created using the \texttt{designmatch} \texttt{R} package. This dataset was created such that all standardized covariate mean differences are below 0.1. This dataset contains 240 subjects, 120 of whom are in treatment and 120 are in control. Thus, compared to the full Lalonde dataset, this dataset contains a subset of the treatment and control groups.
\end{itemize}

Figure \ref{fig:lalondeLovePlot} displays a Love plot of the standardized covariate mean differences for these three datasets. To create this plot, we first defined the covariate matrix and treatment indicator for these three datasets:
\begin{verbatim}
  #obtain the covariates for these datasets
  X.lalonde = subset(lalonde, select = -c(treat))
  X.matched.ps = subset(lalonde.matched.ps, select = -c(treat,subclass))
  X.matched.card = subset(lalonde.matched.card, select = -c(treat,subclass))
  #the treatment indicator is
  indicator.lalonde = lalonde$treat
  indicator.matched.ps = lalonde.matched.ps$treat
  indicator.matched.card = lalonde.matched.card$treat
\end{verbatim}
Then, one can use the \texttt{getStandardizedCovMeanDiffs()} function within \texttt{randChecks} to define the standardized covariate mean differences:
\begin{verbatim}
  meanDiffs.lalonde = getStandardizedCovMeanDiffs(X = X.lalonde,
    indicator = indicator.lalonde)
  meanDiffs.matched.ps = getStandardizedCovMeanDiffs(X = X.matched.ps,
    indicator = indicator.matched.ps)
  meanDiffs.matched.card = getStandardizedCovMeanDiffs(X = X.matched.card,
    indicator = indicator.matched.card)
\end{verbatim}
This allows us to create the Love plot in Figure \ref{fig:lalondeLovePlot}.

Meanwhile, Figure \ref{fig:lalondePermPlot} displays the Love plot for the 1:1 propensity score matched dataset, along with quantiles for the standardized covariate mean differences under complete randomization (i.e., permutations of the treatment indicator). This figure acts as a visualization of our randomization test using the standardized covariate mean differences as a test statistic; it was generated using the \texttt{lovePlot()} function within \texttt{randChecks}:
\begin{verbatim}
  lovePlot(X = X.matched.ps, indicator = indicator.matched.ps,
    permQuantiles = TRUE,
    perms = 1000,
    assignment = "complete")
\end{verbatim}
The argument \texttt{permQuantiles = TRUE} states that the quantiles across permutations should be plotted; \texttt{perms = 1000} states that 1,000 permutations should be used; \texttt{assignment = "complete"} states that the permutations should be under complete randomization.

Finally, Figure \ref{fig:lalondeMDPlot} displays the distribution of the Mahalanobis distance under complete randomization, paired randomization, and constrained randomization (where the standardized covariate mean differences are constrained to be less than 0.1) for the 1:1 cardinality matched dataset. This figure acts as a visualization of our randomization test using the Mahalanobis distance as a test statistic. To assess paired randomization, we needed to define the pair labeling (known as a \texttt{subclass} within the \texttt{randChecks} \texttt{R} package):
\begin{verbatim}
  subclass.matched.card = lalonde.matched.card$subclass
\end{verbatim}
Then, Figure \ref{fig:lalondeMDPlot} was generated using the \texttt{asIfRandPlot()} function within \texttt{randChecks}:
\begin{verbatim}
asIfRandPlot(X = X.matched.card, indicator = indicator.matched.card,
  assignment = c("complete", "blocked", "constrained diffs"),
  subclass = subclass.matched.card,
  perms = 1000,
  threshold = 0.1)
\end{verbatim}
This function was also used to generate Figure \ref{fig:keeleMDPlot} in the real data application (Section \ref{s:realDataAnalysis}). Furthermore, our randomization test results in Sections \ref{ss:graphicalDiagnostic}, \ref{s:simulations}, and \ref{s:realDataAnalysis} were generated using the \texttt{asIfRandTest()} function within \texttt{randChecks}, which has the same arguments as the \texttt{asIfRandPlot()} function.

\section{Data Generating Process for Simulations} \label{s:appendixB}

Here we provide details about the data generating process from the simulation study in Section \ref{s:simulations}. We followed the simulation setup of \cite{austin2009some} and \cite{resa2016evaluation}. We generated 1,000 datasets, where each dataset contained $N_T = 250$ treated subjects and $N_C = 500$ control subjects. Each subject has eight covariates, generated as such:
\begin{equation}
\begin{aligned}
  (x_{i1}, x_{i2}, x_{i3}, x_{i4}) &\sim \mathcal{N}_4 \left( W_i \begin{pmatrix}
    0.2 \\
    0.2 \\
    0.5 \\
    0.5
  \end{pmatrix}, \label{eqn:dataGeneratingModel}
  \begin{pmatrix}
     1 & 0 & 0 & 0 \\
     0 & 1 & 0 & 0 \\
     0 & 0 & 1 & 0 \\
     0 & 0 & 0 & 1
   \end{pmatrix} \right) \\
   (x_{i5}, x_{i6}) &\sim \text{Bern} \left( 0.1 + 0.068W_i \right) \\
   (x_{i7}, x_{i8}) &\sim \text{Bern} \left( 0.4 + 0.242W_i \right)
\end{aligned}
\end{equation}
where $W_i = 1$ if subject $i$ is treated and 0 otherwise. This is similar to ``Scenario 1'' of \cite{resa2016evaluation}; the other two scenarios consider heterogeneous variances and collinearity between the treatment and control groups, and we differ to their work for the performance of matching under those scenarios. The above covariates are generated such that the true standardized difference in means---which is $\frac{\mu_T - \mu_C}{\sqrt{\frac{\sigma^2_T + \sigma^2_C}{2}}}$ for Normal random variables\footnote{Here, $\mu_T$ and $\mu_C$ are the population-level means for the treatment and control groups, respectively, and $\sigma^2_T$ and $\sigma^2_C$ are analogously defined for the variances.} and $\frac{p_T - p_C}{\frac{\sqrt{p_T(1-p_T) + p_C(1 - p_C)}}{2}}$ for Bernoulli random variables\footnote{Here, $p_T$ and $p_C$ are the population-level proportions for the treatment and control groups, respectively.} \citep{rosenbaum1985constructing}---is 0.2 for $\mathbf{x}_1, \mathbf{x}_2, \mathbf{x}_5$, and $\mathbf{x}_6$, and 0.5 for $\mathbf{x}_3, \mathbf{x}_4, \mathbf{x}_7$, and $\mathbf{x}_8$.

After the covariates were generated, three outcomes were generated for each subject:
\begin{align}
  y_{i1} = f_1(x_i) + W_i + \epsilon_i \\
  y_{i2} = f_2(x_i) + W_i + \epsilon_i \\
  y_{i3} = f_3(x_i) + W_i + \epsilon_i
\end{align}
where $\epsilon \stackrel{iid}{\sim} N(0,4)$. Thus, the outcomes are generated noisily around the mean functions $f_1, f_2$, and $f_3$, with an additive treatment effect of one. The mean functions are:
\begin{align}
  f_1(x_i) &= 3.5 x_{i1} + 4.5 x_{i3} + 1.5 x_{i5} + 2.5 x_{i7} \\
  f_2(x_i) &= f_1(x_i) + 2.5 \text{sign}(x_{i1})\sqrt{|x_{i1}|} + 5.5 x_{i3}^2 \\
  f_3(x_i) &= f_2(x_i) + 2.5x_{i3} x_{i7} - 4.5|x_{i1} x_{i3}^3|
\end{align}
Thus, the outcomes $\mathbf{y}_1, \mathbf{y}_2$, and $\mathbf{y}_3$ are ordered in terms of increasing complexity. Furthermore, only the odd-numbered covariates are included in the outcome, in order to mimic the fact that not all available covariates necessarily affect the outcome in practice \citep{resa2016evaluation}.

\vskip 0.2in

\bibliography{randomizationBalanceTestBib}

\bibliographystyle{apa-good}

\end{document}